\documentclass[12pt,final]{l4dc2021}

\usepackage[T1]{fontenc}
\usepackage{amsfonts,dsfont} 
\usepackage[latin5]{inputenc}
\usepackage{graphicx,times,latexsym}
\usepackage{psfrag}
\usepackage{setspace}
\usepackage{algorithmicx,algorithm,algpseudocode}
\usepackage{comment}
\usepackage{color}
\usepackage{url}
\usepackage{booktabs}

\newenvironment{proof}{\paragraph{Proof:}}{\hfill$\square$}
\newtheorem{defn}{Definition}
\newtheorem{exmp}{Example}

\newtheorem{Proposition}{Proposition}

\newcommand{\oprocendsymbol}{\hbox{$\bullet$}}
\newcommand{\oprocend}{\relax\ifmmode\else\unskip\hfill\fi\oprocendsymbol}

\newcommand{\RN}[1]{%
  \textup{\uppercase\expandafter{\romannumeral#1}}%
} 
\providecommand{\figref}{Fig.\,\ref}

\newcommand{\nats}{\mathbb{N}}
\providecommand{\hA}{\hat{A}}
\providecommand{\htheta}{\hat{\Theta}}
\providecommand{\tW}{{\tilde{W}}}

\providecommand{\iid}{\mathrm{i.i.d.}}

\providecommand{\Pdec}{P_{\mathrm{Dec}}}

\providecommand{\Pfa}{P_\mathrm{FA}}

\newcommand{\norm}[1]{\| #1 \|}

\allowdisplaybreaks

\graphicspath{{epsfiles/}}

\title[Learning based attacks in Cyber Physical Systems]{
\LARGE \bf  Learning-based attacks in Cyber-Physical Systems: \\ Exploration, Detection, and Control Cost trade-offs
}
\author{%
  \Name{Anshuka Rangi}$^{{\color{blue}*}}$ \Email{arangi@ucsd.edu}
  \\
  \addr Department of Electrical and Computer Engineering, University of California San Diego,
  CA 92093
  \AND
  \Name{Mohammad Javad Khojasteh}\thanks{indicates equal contribution.} \Email{mkhojast@mit.edu}\\
  \addr Laboratory for Information and Decision Systems, Massachusetts Institute of Technology,
   MA, 02139
  \AND
  \Name{Massimo Franceschetti} \Email{mfranceschetti@ucsd.edu}
  \\
  \addr Department of Electrical and Computer Engineering, University of California San Diego,
  CA 92093
}

\begin{document}
\maketitle
\begin{abstract}%
 We study the problem of learning-based attacks in  linear   systems, where the communication channel between the controller and the plant can be hijacked by a malicious attacker. 
 We assume the attacker learns the dynamics of the system from observations, then  overrides the controller's actuation signal, while mimicking legitimate operation by providing fictitious feedback about the sensor readings  to the controller.  
On the other hand, the controller is  on a lookout to detect the presence of the attacker and tries to  enhance the detection performance by carefully crafting its control signals. 
 We study the trade-offs between    the information acquired by the attacker from observations,  the detection capabilities of the controller, and the control cost. Specifically, we provide  tight upper and lower bounds on the expected $\epsilon$-deception time, namely the time required by the controller to make a decision regarding the presence of an attacker with confidence at least $(1-\epsilon\log(1/\epsilon))$.  We then show a probabilistic lower bound  on the time that must be spent by the attacker learning the system, in order for the controller to have a given expected $\epsilon$-deception time. We show that this bound is also order optimal, in the sense that 
   if the attacker satisfies it, then there exists a learning algorithm with the given order expected deception time. Finally, we show a lower bound on the expected energy expenditure required to guarantee detection with confidence at least $1-\epsilon \log(1/\epsilon)$.
\end{abstract}

\begin{keywords}%
Data poisoning attack, Man in the middle attack, Cyber physical system
\end{keywords}

\section{Introduction}
\label{s:intro}
Attacks directed to  Cyber-Physical Systems (CPS)
  can have catastrophic consequences 
ranging from hampering the economy through financial scams,  to possible losses of human lives through hijacking autonomous vehicles and drones, see~\cite{pasqualetti2013attack,shoukry2018smt,hoehn2016detection}. In this framework, two important problems arise: understanding of the regime where the system  can be attacked, and designing ways to mitigate these attacks and render the system secure, see \cite{ma2019policy,zhang2020online,jun2018adversarial,zhan2020preventing,chen2019optimal,vemprala2020adversarial,ferdowsi2018deep,mao2020novel,khojasteh2019authentication,khojasteh2018authentication,rangi2021secure}.  
Techniques developed to secure CPS include  watermarking, moving target and baiting, and typically require either a loss of performance, or additional resources available at the controller, see \cite{Kumar:DynamicWatermarki,MoSinopoli:Magzine, kanellopoulos2019moving,flamholz2019baiting}.

In this paper, we focus on the former aspect of the problem, namely understanding the regime under which the system can be attacked.
We focus on linear plants and on an important and widely used class of attacks  based on the ``man-in-the-middle"  (MITM) technique. In this case, 
the attacker   takes over the physical plant's control and feedback signals, and acts as a malicious controller for the plant and fictitious plant for the controller. By doing so, it overrides the control signals 
with malicious inputs aimed at destroying the plant; and it overrides the feedback signals to 
the controller, trying to mimic the safe and legitimate operation of the system.  
In learning based MITM attack, we   assume that  the attacker has full access to both sensor and  control signals, but the plant dynamics are unknown to the attacker. 
Thus, the attacker needs to learn about the plant in order to being able to generate the fictitious signals to the controller that allow the attacker to  remain  undetected for the time needed to cause harm.
On the other hand, the controller has perfect (or nearly perfect) knowledge of 
the system dynamics  and  is actively looking out for an anomalous behaviour in the feedback signals from the plant.  This assumed information pattern    is justified, since the controller is typically tuned in much longer than the attacker, and has knowledge of the system dynamics to a far greater precision than the attacker.
Following the detection of the attacker, the controller can shut the plant down, or  switch to a ``safe'' mode where the system is secured using additional resources, and the attacker is prevented  from causing additional "harm" to the plant, see
\cite{
dibaji2019systems,
weerakkody2019resilient,teixeira2015secure,hashemi2020gain}. 

We consider a learning-based MITM attack   that evolves in two phases: \textit{exploration and exploitation}. In the exploration phase, the attacker  observes the plant state and control inputs, and  learns the plant dynamics. In the  exploitation phase, the attacker hijacks the plant, and utilizes the learned estimate to feed the fictitious feedback signals to the controller. During this phase, the attacker may also refine its estimate by continuing to learn. Within this context, our results are as follows: first, we provide a lower bound on the expected $\epsilon$-deception time, namely the  time required by the controller to make a decision regarding the presence of an attacker with confidence at least $1-\epsilon\log(1/\epsilon)$. This bound is expressed in terms of the parameters of the attacker's learning algorithm and the controller's  strategy. Second, we show that there exists a learning-based attack and a detection strategy such that a matching upper bound on the expected $\epsilon$-deception time is obtained. 
We then show that for a wide range of learning algorithms, if 
 the expected $\epsilon$-deception time is at least of duration $D$, 
then the duration of the exploration phase of the attacker must be at least  $\Omega({D/\log(1/\epsilon)})$,  as $\epsilon\to 0$. We establish that this bound is also  order-optimal since there exists a learning algorithm such that if the duration of the exploration phase is $O({D/\log(1/\epsilon)})$ as $\epsilon\to 0$, then the expected $\epsilon$-deception time is at least $D$. 
Finally,  we   show that if the controller wants to detect the attacker in at most $D$ duration  with confidence at least $1-\epsilon\log(1/\epsilon)$, then the expected energy expenditure on the control signal must be at least of order $\Omega({D}/\log(1/\epsilon))$,  as $\epsilon\to 0$. 

All proofs
are available in appendix. 

\section{Related Work} 
There is a wide range of recent research  on learning-based control for linear systems ~\cite{Dean2019,sarkar2019near,berkenkamp2017safe,fisac2018general,khojasteh2019probabilistic,vrabie2009adaptive,jiang2012computational,cheng2020safe,fan2020deep,lederer2019uniform,buisson2020actively}. In these works, learning algorithms are proposed to design controllers in the presence of uncertainty. In contrast, in our setting  we assume that the controller has full knowledge of the system dynamics, while the attacker may take advantage of these algorithms. Thus, our focus is not on the optimal control design given the available data, but rather on the trade-offs between   the  attacker's learning capability, the  controller's  detection strategy, and the control cost.

The MITM attack has been  extensively  studied in control systems for two special cases, namely, the replay attack and the statistical duplicate attack. The detection of replay attacks has been studied in \cite{mo2014detecting,MoSinopoli:Magzine,miao2013stochastic}, and ways to mitigate these attacks have been studied in \cite{zhu2014performance}.  Likewise, the ways to detect and mitigate statistical duplicate attacks has been studied in \cite{Kumar:DynamicWatermarki,smith2015covert,porter2020detecting}. 
These works do not consider   learning-enabled attackers, and analyze the performance of the controller for only a specific detection strategy. In contrast, we investigate   learning-enabled attacks, and present   trade-offs between   the  attacker's learning capability through observations,  the controller detection strategy,  and the control cost. 
Learning based attacks have been recently considered in \cite{khojasteh2019authentication,khojasteh2018authentication, ziemann2020parameter}. In \cite{khojasteh2019authentication,khojasteh2018authentication}, a variance based detection strategy has been investigated to present bounds on  the probabilities of detection (or false alarm) of the attacker. 
In~\cite{ziemann2020parameter},   an optimization-based controller is proposed that has the additional capability of injecting noise to  interfere with the learning process of the attacker. Here, we consider a  wider  class of learning-based attacks and detection strategies, and provide tight trade-offs for these attacks. 


Multiple variants of MITM attacks are studied in Reinforcement Learning (RL). In \cite{rakhsha2020policy}, the work studies the MITM attacks  under the assumption that the attacker has perfect knowledge of the underlying MDP. The results are further extended to the setting where attacker has no knowledge of the underlying MDP \cite{rakhsha2021reward}. This is analogues to studying learning based attacks in RL where the attacker eavesdrops on the actions performed by the learner and manipulates the feedback from the environment. 
In \cite{zhang2020adaptive}, the work studies the feasibility of MITM attack under the constraint on the amount of contamination introduced by the attacker in the feedback signal. 
The relationship between the problem of designing optimal MITM attack in RL and the problem of designing optimal control is discussed in \cite{zhu2018optimal}. 
Finally, the learning based MITM attacks are also an active area of research in the Multi-Armed Bandits (MAB), see \cite{jun2018adversarial,ma2018data,bogunovic2020stochastic,rangi2021secure}. In the same spirit of our work, these works study the feasibility of the attacks,  and provide  bounds on the amount of contamination needed by the attacker to achieve its objective. However, these works do not consider the possibility of the detection of the attacker. In this work, we focus on understanding the regime where the system can be attacked without the detection of the attacker.

  \section{Problem Setup}
\label{sec:formulate}

\begin{figure}
  \centering
  \subfigure[Exploration Phase\label{fig:sn1}]{\includegraphics[scale=0.45]{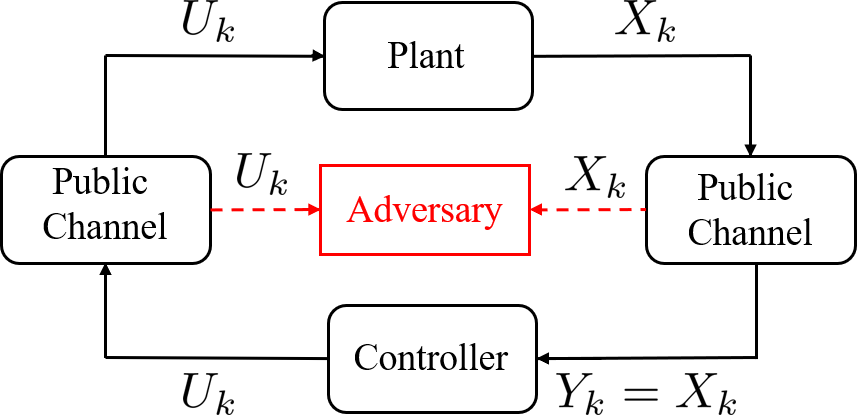}} \qquad\qquad\qquad
 \subfigure[Exploitation Phase\label{fig:sn2}]{\includegraphics[scale=0.4]{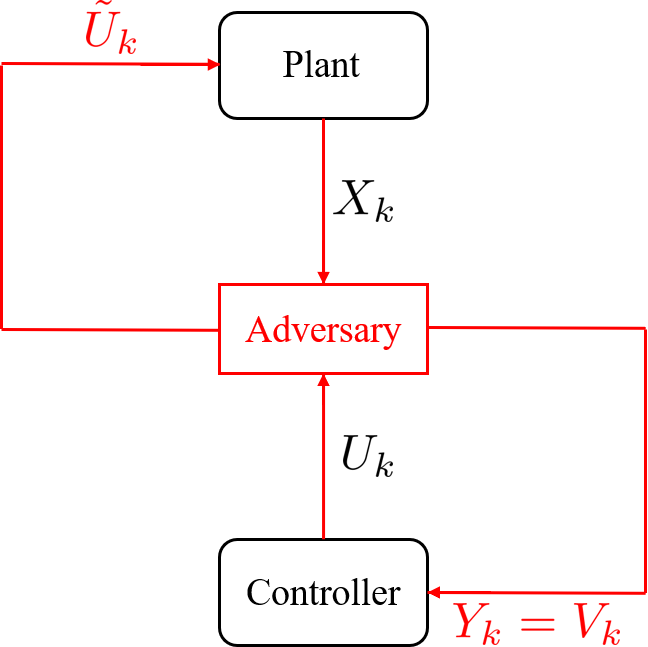}}
  \caption{System model during the two attack phases.}\label{fig:sns}
\end{figure}

We consider the networked control system depicted in \figref{fig:sn1} and \figref{fig:sn2}, where the plant dynamics are described by a discrete-time and linear time-invariant (LTI) system, namely  at time $k \in \nats$, we have
	\begin{align}
 	\label{eq:plant}
   		X_{k+1}=AX_k+U_k+W_k,
 	\end{align} 
    where $X_k$,  $U_k$, $W_k$ are vectors of dimension $M\times 1$ representing the plant state,  control input, and plant disturbance respectively, and $A$ is a matrix of dimension $M\times M$, representing the open-loop gain of the plant. At time $k$, the controller observes the feedback signal $Y_k$
and generates a control signal $U_k$ as a function of $Y_{1:k}=\{Y_{1},\ldots, Y_{k}\}$.
The initial state $X_0$ is known to both the controller and the attacker, and is 
independent of the disturbance sequence $\{W_k\}_{k=1}^{\infty}$, where $W_k$ is  i.i.d. Gaussian noise $\mathcal{N}(0,\sigma^2I_M)$  with PDF known to both the parties, and $I_M$ is the identity matrix of dimension $M\times M$. Our results can also be extended to the scenario where the PDF of the noise known to the attacker is different from the actual PDF of the noise (or PDF known to the controller).  
With a slight loss of generality, we assume that $U_0 = W_0 = 0$ for analysis. 
   
    The controller attempts to detect  the presence of the attacker based on the observations $Y_{1:k}$.  When the controller detects an attack, it shuts the system down and prevents the attacker from causing further ``damage'' to the plant. The controller is aware of the plant dynamics in ~\eqref{eq:plant}, and knows the  gain $A$. This is justified because one can assume that  the controller is tuned  to the plant for a long duration and thus has   knowledge of $A$ to a  great precision. On the other hand, the attacker only knows the form of the state evolution equation ~\eqref{eq:plant}, but does not know the  gain matrix $A$.




\section{Learning based Attacks}
\label{ss:model:integrity_attack}
We consider learning based attacks that evolve in two phases.

\textit{Phase 1: Exploration.}  Let $L$ be the duration of the exploration phase.  
For all $k\leq L$, as illustrated in \figref{fig:sn1}, the attacker passively 
eavesdrops on the control input $U_{k}$ and the plant state $Y_k=X_{k}$ with the objective of learning the open loop gain of the plant. 
We let $\hA_{k}$ be the attacker's estimate of $A$ at time step $k$. The duration $L$ can be considered as   the cost incurred by the attacker, since its actions are limited to eavesdropping during this phase. 

\textit{Phase 2: Exploitation.}
The exploration phase is followed by the exploitation phase. For all $k\geq L+1$, as illustrated in \figref{fig:sn2}, the attacker  hijacks the system  and feeds a malicious control signal $\tilde{U}_k$ to the plant in order to destroy the plant. Additionally, the attacker may continue to learn about $A$,  and utilizes its estimate $\hA_{k}$ to design a fictitious feedback signal $Y_k=V_k$ in \figref{fig:sn2} to deceive the controller, namely
\begin{align}
\label{mimicmodel}
	V_{k+1} &= \hA_{k} V_k + U_k + \tW_k,
\end{align}
where for all $k \geq L+1$, $\tW_k$ are i.i.d.\ with $f_{\tW}=f_{W}=\mathcal{N}(0,\sigma^2I_M)$
.  Let $R$ denote an attack strategy whose feedback signal satisfies \eqref{mimicmodel}.
Thus, for all $L>0$, our class of learning based attacks is
\begin{equation}\label{eq:classOfAttack}
\begin{split}
\mathcal{A}(L)&= \{R: \mbox{for all $k\leq L$, } Y_k=X_k \mbox{ and for all $k\geq L+1$, }Y_k=V_k \}.
\end{split}
\end{equation} 
 Note that in the class $\mathcal{A}(L)$, the learning of $A$ may or may not continue during the exploitation phase. Additionally, the attacker may use different learning algorithms in the two phases. 

If the attacker learns   $A$ perfectly, i.e. $\hat{A}_{k}=A$, then \eqref{mimicmodel} will perfectly mimic the plant behavior, making it impossible for the controller to detect the attacker. 
Otherwise, the controller can attempt to detect the presence of the attacker by testing for statistical deviations from the typical behavior in ~\eqref{eq:plant}. The following example illustrates this point. 

\begin{exmp}
Let 
$R^*\in \mathcal{A}(L)$ be an attack whose learning is only limited to the exploration phase, namely $\hat{A}_{k}=\hat{A}_{L}$ for all $k\geq L+1$. Also, let $\norm{\cdot}_{op}$ be the
\textit{operator}
norm induced by the Euclidean norm $\norm{\cdot}_2$ when applied to a matrix.
In the exploration phase there is no interference from the attacker and  for all $k\leq L$, the  observation $Y_k=X_k$ satisfies
\begin{align}
\label{L2Lshouldsatf}
	Y_{k+1} - A Y_k - U_k= W_k\sim ~\iid~ f_W.
\end{align}
In the exploitation phase,  for all $k\geq L+1$,  the controller observation $Y_k=V_k$ satisfies
\begin{equation}\label{eq:substract4test:reduced}
 	V_{k+1}-AV_k-U_k
  = V_{k+1} - AV_k + \hA_L V_k - \hA_L V_k-U_k
 = \tW_k + \left( \hA_{L} - A \right) V_k,
\end{equation}
where 
\eqref{eq:substract4test:reduced} 
follows from \eqref{L2Lshouldsatf}. 
Since $\tW_k$ and $W_{k}$  have the same distribution and $\norm{Ax}_2\leq \norm{A}_{op}\norm{x}_2 $ holds, the controller can test the statistical deviation of \eqref{L2Lshouldsatf} from  \eqref{eq:substract4test:reduced}. In this case,
the detection of the attack is controlled by two factors: the estimation error $\norm{\hA_{L} - A}_{op}$ and the fictitious signal $V_{k}$.

At the controller's side, the detection becomes easier when  the error $\norm{\hA_{L} - A}_{op}$ increases.  Thus, at the attacker's side  it is desirable to reduce the error $\norm{\hA_{L} - A}_{op}$. This can be done by increasing the duration $L$, and incurring an additional learning cost.

The detection is also easier  if the energy of the fictitious signal $V_{k}$ is large.  Since $V_{k}$ is a function of the control signal $U_{k-1}$, it follows that the energy spent by the controller can help in the detection of the attacker.

We then conclude that  the probability of successful detection (or the time required to detect the attacker with a given confidence) should reveal a trade-off between the duration $L$ of the exploration phase (or the estimation error $\norm{\hA_{L} - A}_{op}$), and the energy of the fictitious signal (or  of the control signal).
In this paper  
  we quantify both upper bound and lower bound on this trade-off.
  \end{exmp}
 \subsection{Performance Measures}
\begin{defn}
The decision time $\tau$ is the time  at which the controller makes a decision regarding the presence or absence of the attacker. 
\end{defn}

\begin{defn} The probability of deception   is the  probability of the attacker deceiving the controller and remaining undetected at the decision time $\tau$, namely $\Pdec^{\tau} \triangleq 				\mathbb{P}
       ( \htheta_\tau=0 | \Theta_\tau=1 )$,
where $\hat{\Theta}_{\tau}$ denotes the decision of the controller at the decision time $\tau$, and the hijack indicator $\Theta_k$ at time $k$ is
\begin{align}
\label{theta1s}
	\Theta_k \triangleq
    \begin{cases}
        0, & \forall j \le k:\ Y_j = X_j \,;
	 \\ 1, & \textrm{otherwise} .
    \end{cases}
\end{align}
Likewise, the probability of false alarm is the probability of detecting the attacker when it is not present at the decision time $\tau$, namely $\Pfa^{\tau} \triangleq 
        \mathbb{P}
        ( \htheta_\tau=1 | \Theta_\tau=0 ).$
\end{defn}
In the class  $\mathcal{A}(L)$ in \eqref{eq:classOfAttack}, for all  $k\leq L$, we have that  $\Theta_k=0$ (exploration phase); and  for all $k\geq L+1$, we have
 $\Theta_k=1$   (exploitation phase).  
 \begin{defn} \label{def:DeceptionTime}
For all attacks in the class  $\mathcal{A}(L)$ and  $0<\epsilon<1$, the $\epsilon$-deception time $T(\epsilon)$ is the  time required by the controller to make a decision,  with 
$\Pdec^{\tau}\leq \epsilon \log(1/\epsilon)$, where $\tau=L+T(\epsilon)+1$.
 \end{defn}
 Thus, $T(\epsilon)$ is the largest possible duration during which the attacker can deceive the controller, and remain  undetected with confidence at least $1-\epsilon\log(1/\epsilon)$, namely  for all $L+1\leq k\leq T(\epsilon)+L$, we have 
 \begin{equation}\label{eq:ProbDetect}
   \mathbb{P}(\htheta_k = \Theta_k|\Theta_k=1)=\mathbb{P}(\htheta_k = 1|\Theta_k=1)< 1-\epsilon\log(1/\epsilon). 
\end{equation}

\begin{defn}
For all $n>L$, the expected deception cost of the attacker until time $n$ is defined as 
\begin{align}
\label{cnequation1}
    C(n)\triangleq \frac{1}{n} \mathbb{E}\bigg[\sum_{k=L+1}^{n}\frac{V_{k-1}^{T}(\Hat{A}_{k-1}-A)^T(\Hat{A}_{k-1}-A)V_{k-1}}{2\sigma^2}\bigg]. 
\end{align}

\end{defn}
\subsection{Main results}

 
   We start with defining a non-divergent learning algorithm.

\begin{defn}
A learning algorithm is non-divergent if its estimation error is non-increasing in the duration of the learning, namely for all $k_2>k_1$, we have $\|\Hat{A}_{k_2}-A\|_{op}\leq \|\Hat{A}_{k_1}-A\|_{op}.$
\end{defn}
We introduce the following notation. Let  $p_{0}(y_{1:\tau})$ be the conditional probability of $y_{1:\tau}$ given the attacker did not hijack the system, namely $\Theta_{1}=\ldots\Theta_{L}=\Theta_{L+1}=\ldots\Theta_{\tau}=0$, where  $y_{1:\tau}$ denotes the realization of the random variables $Y_{1},\ldots , Y_{\tau}$. Likewise, let $p_{1}(y_{1:\tau})$ be the conditional probability of $y_{1:\tau}$ given the attacker has hijacked the system, namely $\Theta_1=\ldots=\Theta_{L}=0$ and  $\Theta_{L+1}=\ldots\Theta_{\tau}=1$. The following proposition  characterises the KL divergence $D(p_{1}(Y_{1:\tau})||p_{0}(Y_{1:\tau}))$ between $p_{1}(Y_{1:\tau})$ and $p_{0}(Y_{1:\tau})$, 
and is useful to derive our main results.
\begin{Proposition}\label{lemma:KLDivergence}
For all attacks in the class $\mathcal{A}(L)$  and 
$n>L$, the cumulative KL divergence is
\begin{equation}
    D(p_{1}(Y_{1:n})||p_{0}(Y_{1:n}))=n C(n).
\end{equation}
\end{Proposition}
The KL divergence between the distributions $p_0$ and $p_1$ is characterized by $C(n)$, and is the key quantity to establish both the lower bound and the upper bound on  $T(\epsilon)$. 
If the PDF of the noise known to the attacker is different from the actual PDF of the noise (or the PDF known to the controller), Proposition \ref{lemma:KLDivergence} can be modified to include this discrepancy, and  an additional non-negative term would be added to $C(n)$. The bounds on $T(\epsilon)$ will follow along the same lines. 

The following theorem presents a lower bound on $\mathbb{E}[T(\epsilon)]$ that holds for any detection strategy. The bound is expressed in terms of $C(n)$, which depends   on the attacker's learning algorithm, the fictitious signal and the control signal in \eqref{mimicmodel}.
 
\begin{theorem} \label{thm:timeErrorEnergy}
For all attacks in $\mathcal{A}(L)$ and $\tau>L$, if
\begin{align}\label{eq:Error1}
    \Pdec^{\tau}=O(|\epsilon\log\epsilon|) \mbox{ and } P_{FA}^{\tau}=O(|\epsilon\log\epsilon|), \mbox{ as } \epsilon\to 0,
\end{align}
then 
the deception time $T(\epsilon)=\tau-L-1$  is
\begin{align}\label{eq:decisionTime}
    \mathbb{E}[T(\epsilon)]\geq \frac{\log(1/\epsilon)}{ C(n_0)} +o(\log(1/\epsilon)) \;\; \mbox{ as } \epsilon\to 0,
\end{align}
 where  $n_{0}=\max \left\{n> L: n C(n)< \log(1/\epsilon)\right\}.$
\end{theorem}
It follows  that for any detection strategy with probability of error $O(|\epsilon \log\epsilon|)$, the expected $\epsilon$-deception time is at least 
$\Omega\left(\log(1/\epsilon)/C(n_0)\right)$. 
The next theorem establishes that the lower bound   in Theorem \ref{thm:timeErrorEnergy} is tight.
\begin{theorem}\label{thm:ExistenceOfTest}
 There exists an attack in $ \mathcal{A}(L)$ and a detection strategy such that at  $\tau>L$, we have
\begin{align}\label{eq:testerror1}
    \Pdec^{\tau}=O(\epsilon) \mbox{ and } P_{FA}^{\tau}=O(\epsilon), \mbox{ as } \epsilon\to 0,
\end{align}
and the   deception time $T(\epsilon)=\tau-L-1$ is 
\begin{align}\label{eq:decisionTime2}
    \mathbb{E}[T(\epsilon)]\leq \frac{\log(1/\epsilon)}{C(n_{0}+1)}+o(\log(1/\epsilon)), \;\; \mbox{ as } \epsilon\to 0.
\end{align}
\end{theorem}
In Theorems \ref{thm:timeErrorEnergy} and \ref{thm:ExistenceOfTest}, as $\epsilon\to 0$, we have that $C(n_0) \to C(n_0+1)$, and $|\epsilon| \le |\epsilon \log \epsilon|$. 
Thus, the lower bound and the upper bound in Theorems \ref{thm:timeErrorEnergy} and \ref{thm:ExistenceOfTest} are tight.  
It turns out that the attack achieving the upper bound on $\mathbb{E}[T(\epsilon)]$ in  Theorem  \ref{thm:ExistenceOfTest}   learns about $A$ in the exploration phase only, and focuses on destabilizing the system in the exploitation phase. 
The corresponding detection strategy is a classic sequential probability ratio  test (\cite{wald1948optimum}), which computes the ratio of the posterior probability of the two hypotheses, namely the attacker is present or absent, and makes a decision when this ratio crosses the threshold $\log(1/\epsilon)$. While this strategy has been previously studied under the assumption that the samples $y_{1:n}$ are identically and independently distributed (i.i.d) (\cite{chernoff1959sequential,rangi2018decentralized,rangi2018consensus,rangi2020distributed}), here we extend the analysis to the samples dependent on both
the control input  and the state   of the feedback signal at the controller.

We point out that to extend these results to non-linear systems, a key step  would be finding an analogue of Proposition \ref{lemma:KLDivergence} in a  non-linear setting. This proposition relates the KL divergence to the expected deception cost $C(n)$, which is a function of the fictitious signal and the error in the estimation of $A$. For non-linear systems, an equivalent relationship needs to be derived between the KL divergence, the fictitious signal and the error in the estimation of non-linear system dynamics. The  proof of the Theorems \ref{thm:timeErrorEnergy} and \ref{thm:ExistenceOfTest} can then be obtained using a similar argument,   given an analogue of Proposition \ref{lemma:KLDivergence} for non-linear systems. 

Next, we derive some useful implications of Theorems \ref{thm:timeErrorEnergy} and \ref{thm:ExistenceOfTest}.
For simplicity of presentation, in the following we restrict  the class  of learning algorithms in the exploration phase, although    results can also be  extended to   more general settings. 
\begin{defn}
\label{definitofLeanif}
A learning algorithm is said to be convergent if there exists an $\alpha\geq 1$ such that for all $\eta>0$ and time step $k> 0$, we have 
 \begin{equation}\label{eq:conv}
     \mathbb{P}(\|\hA_{k}-A\|_{op} > \eta)\leq \frac{c}{(\eta^2 k)^{\alpha}}.
 \end{equation}
\end{defn}
It follows that any convergent learning algorithm provides an unbiased estimate of $A$ as the learning time $k\to \infty$, and the operator norm of the estimation error converges to the interval $[-\eta,+\eta]$ at rate $O(1/{(\eta^2 k)^{\alpha}})$. There are many convergent learning algorithms. For example, for scalar systems and  measurable control policy, the Least Squares (LS) algorithm  in ~\cite{rantzer2018concentration} satisfies 
\begin{align}
\label{thebound222}
    \mathbb{P}(|\hA_{k}-a|>\eta) \leq \frac{2}{(1+\eta^2)^{k/2}}.
\end{align}
For the vector case  sufficiently large learning time $k$,  if the control input is $U_k=-\bar{K}X_k$ and $A-\bar{K}$ is a marginally stable matrix, then the LS algorithm in  ~\cite{simchowitz2018learning} satisfies
\begin{align}
\label{smichottee}
    \mathbb{P}(\|\hA_{k}-A\|_{op}>\eta) \le \frac{c_1}{e^{\eta^2 k}},
\end{align}
where $c_1>0$ is a constant. 

The following theorem provides a lower bound on the duration of the exploration phase for the attacker to achieve a given expected $\epsilon$-deception time. 
\begin{theorem}\label{Ltheorem11}
For all    $0<\delta<1$ and $D>0$,  and all attacks in  $\mathcal{A}(L)$ using a convergent learning algorithm  in the exploration phase   and a non-divergent  learning algorithm  in the exploitation phase,
if  $\mathbb{E}[T(\epsilon)]\geq D +o(1)$ as $\epsilon\to 0$, 
then with probability at least $1-\delta$ the following asymptotic inequality holds
\begin{equation}
   L\geq {\frac{D \tilde{C}(n_0)}{\log(1/\epsilon)}}\bigg(\frac{c}{\delta}\bigg)^{1/\alpha}+o\bigg(\frac{1}{\log(1/\epsilon)}\bigg), \mbox{ as } \epsilon\to 0,
\end{equation}
 where $ \tilde{C}(n)=  \mathbb{E}\big[\sum_{k=L+1}^{n}{V_{k-1}^{T}V_{k-1}}\big]/(2\sigma^2 n)$.
\end{theorem}


The following theorem establishes that the lower bound on $L$ in Theorem \ref{Ltheorem11} is order optimal, and a matching order-optimal bound on $L$  holds for the LS algorithm in~\cite{simchowitz2018learning}.

\begin{theorem}\label{thm:algExistence}
For all  $0<\delta<1$ and $D>0$, using the LS  algorithm in ~\cite{simchowitz2018learning} in the exploration phase only, and assuming the control input is $U_k=-\bar{K}X_k$, where $A-\bar{K}$ is a marginally stable matrix, if
 \begin{equation}
     L=D\tilde C(n_0) \log(c_1/\delta) /\log(1/\epsilon) +o(1/\log(1/\epsilon)) \mbox{ as } \epsilon\to 0,
 \end{equation}
 then, with probability  at least $1-\delta$ we have  
\begin{equation}
    \mathbb{E}[T(\epsilon)]\geq D +o(1), \mbox{ as }\epsilon\to 0.
\end{equation}
\end{theorem}
The choice of the control policy can play a crucial role in the reduction of the deception time. However, this can occur at the expense of the energy used to construct the control signal $U_{k}$. The following theorem provides a lower bound on the amount of energy that the controller needs to spend to achieve a desired expected $\epsilon$-deception time. 
\begin{theorem}
\label{controlenergw22}
For all  $D>0$, and for all attacks in $\mathcal{A}(L)$ using a  non-divergent  learning algorithm  in the exploitation phase, if $\mathbb{E}[T(\epsilon)]\leq D+o(1)$ as $\epsilon\to 0$, and for all $k>L$, the control policy satisfies
\begin{align}\label{kkdcondition!}
      \mathbb{E}[V_k^T\hat{A}^T_k \hat{A}_k V_{k}]+\sigma^2+ 2\mathbb{E}[V^T_{k}\hat{A}^T_kU_k]\leq 0,
  \end{align}
  then the expected energy of the control signal  is 
\begin{equation}
    R(n_{0})\geq \frac{2\sigma^2\log(1/\epsilon)}{\|\hA_{L}-A\|_{op}^2 D} + o(\log(1/\epsilon)), \mbox{ as }\epsilon\to 0,
\end{equation}
where $R(n_0)\triangleq  \mathbb{E}\big[\sum_{k=L}^{n_0-1}U_{k-1}^TU_{k-1}\big]/n_0$.
\end{theorem}

Theorem~\ref{controlenergw22} shows that the expected energy of the control signal until a time between $L\leq k\leq n_0$  is inversely proportional to the upper bound $D$ on the deception time. 
Since  $L$ is unknown to the controller, it follows that the controller should maintain a high level of expected signal energy $\mathbb{E}[U_{k}^2]$ at every time instance $k$ to ensure a small deception time. 

\section{Simulations}
\label{sec:simulation1}
In this section, we provide two numerical examples. Although our theoretical results are valid for a large class of learning algorithms  and any detection strategy chosen by the controller, we validate them here using LS algorithm and a covariance detector.

First we start with an example for  scalar system, where we use the empirical performance of a variance-test to illustrate our results. Specifically, at a decision time $\tau$,
the controller
tests the empirical variance for unexpected behaviour over a detection window  $[0, \tau]$, using a confidence interval $2 \gamma > 0$ around the expected variance. More precisely, at decision time $\tau$, $\htheta_\tau= 0$ if
\begin{align}
\label{Test1}
\begin{aligned}
	\frac{1}{\tau} \sum_{k=0}^{\tau} \left[Y_{k+1}-a Y_k-U_k \right]^2 
 \in (\mathrm{Var}[W]-\gamma, \mathrm{Var}[W]+\gamma),
\end{aligned}
\end{align}
otherwise $\htheta_\tau=1$.
In this case, since the system disturbances are i.i.d. Gaussian   $\mathcal{N}(0,\sigma^2)$, using Chebyshev's inequality, we have 
\begin{align}
\label{chebfalsealarm}
    \Pfa^{\tau} \le \frac{\mathrm{Var}[W^2]}{ \gamma^2T}=\frac{3\sigma^4}{\gamma^2T}.
\end{align}
In our simulations, the attacker learns in the exploration phase only, and uses the LS learning algorithm. At the end of the exploration phase, we have 
\begin{align}
\label{learningAlgorithm22}
	\hA_{L} = \frac{\sum_{k=1}^{L-1}(X_{k+1}-U_k)X_k}{\sum_{k=1}^{L-1} X_k^2}.
\end{align}


Our simulation parameters are the following:
$\gamma=0.1$, decision time $\tau=800$, $A= 1.1$, and $\{W_k\}$ are i.i.d. Gaussian $\mathcal{N}(0,1)$.  
Using~\eqref{chebfalsealarm}, the false-alarm rate is negligible for these parameters. 

\begin{figure}
  \centering
  \subfigure[Attacker's success rate versus  $L$\label{fig:mnahayiii2222}]{\includegraphics[scale=0.45]{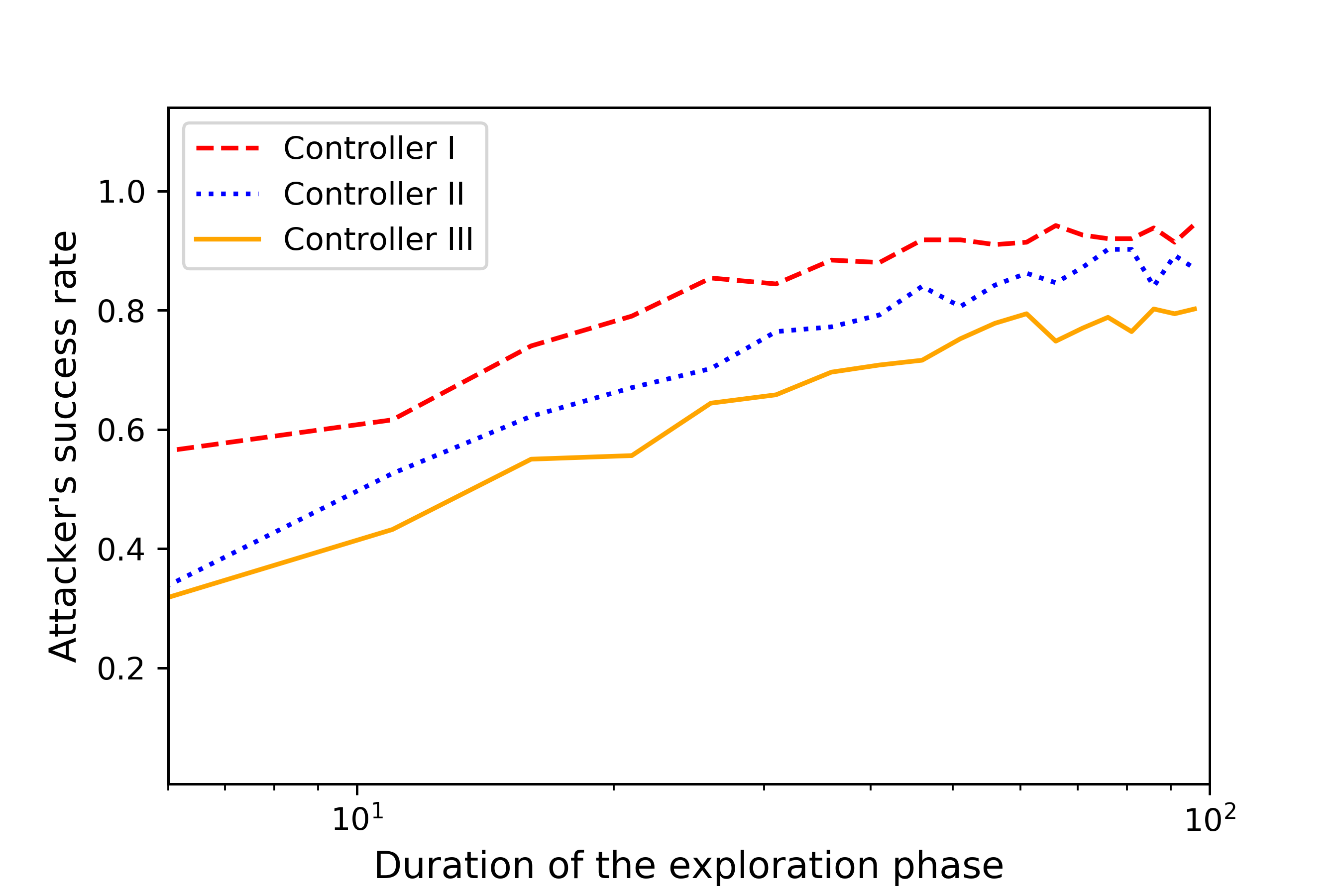}} \quad
 \subfigure[Attacker's success rate versus $\tau$\label{fig:avalsimu}]{\includegraphics[scale=0.15]{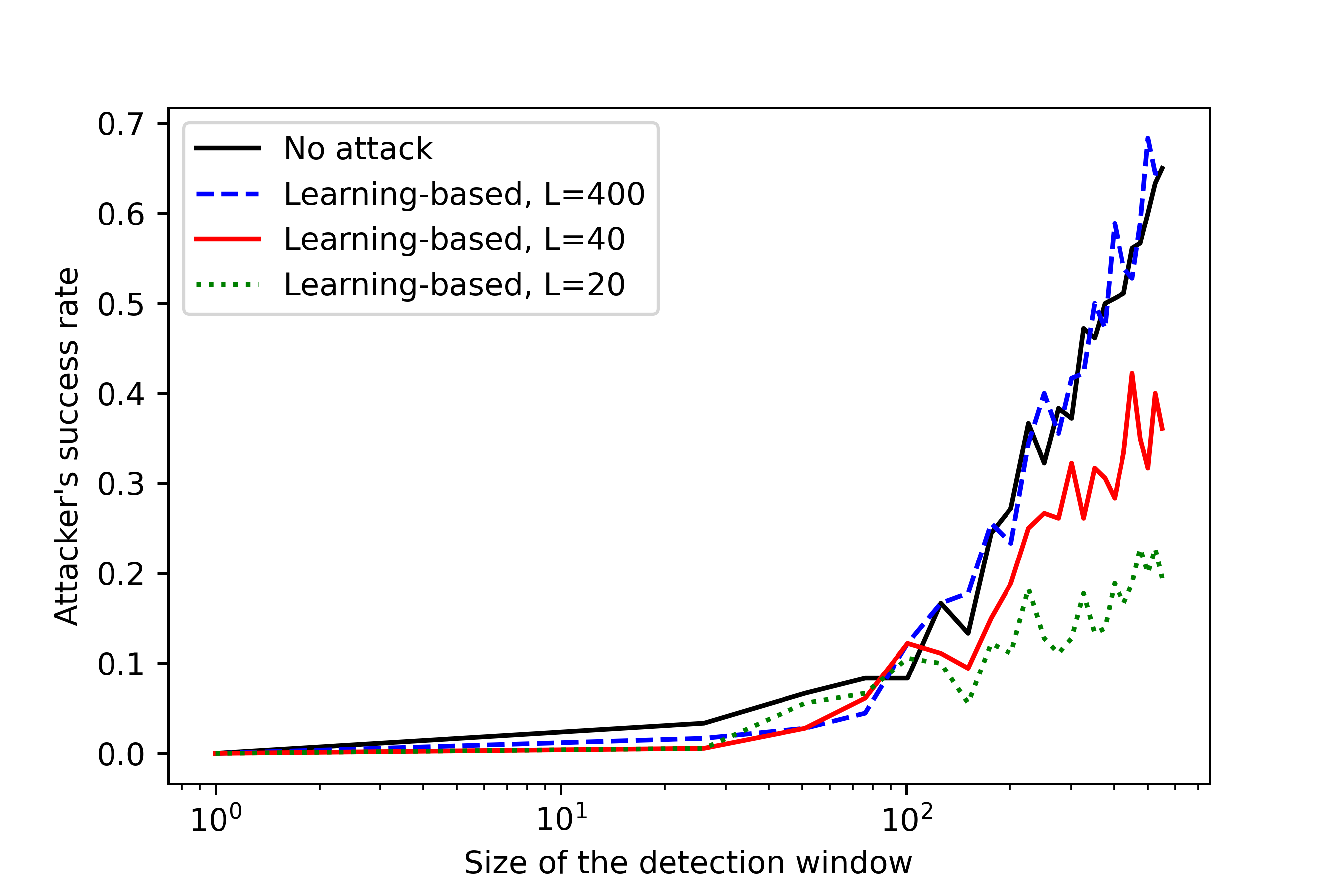}}
  \caption{Simulations Result.}\label{fig:simulation234}
\end{figure}

  
\figref{fig:mnahayiii2222} compares 
the attacker's success rate
as a function of the duration $L$ of the exploration phase
for three different control policies $U_k=-AY_k+\Gamma_k$ such that for all $k$, I) $\Gamma_k=0$,  II) $\Gamma_k$ are i.i.d.\ Gaussian $\mathcal{N}(0,9)$,  III) $\Gamma_k$ are i.i.d.\ Gaussian $\mathcal{N}(0,16)$. As illustrated in~\figref{fig:mnahayiii2222}, the attacker's success rate increases as the duration of exploration phase increases. This is because the attacker's   estimation error $|\hat{A}_{L}-A|$ reduces as $L$ increases, which makes it difficult for the controller to detect the attacker. This is in accordance with the theoretical findings in Theorem \ref{Ltheorem11}.
Also, for a fixed $L$, the attacker's success rate decreases 
as the input control energy increases. The increase in the control energy increases the energy of the fictitious signal which increases the probability of detection, and is in accordance with Theorem \ref{controlenergw22}. 

Next, we provide an example of vector system, and analyze  the empirical performance of the covariance test against the learning-based attack. In vector systems, the  error matrix is 
\begin{align}
	\Delta
	\triangleq 
	\Sigma
    -\frac{1}{\tau} \sum_{k=1}^{\tau} \left[ Y_{k+1}-A Y_k-U_k \right]    \left[ Y_{k+1}-A Y_k-U_k  \right]^\top
 \nonumber
\end{align}
Similar to ~\eqref{Test1}, at decision time $\tau$, we have $\htheta_\tau= 0$ if $\|\Delta\|_{op} \le \gamma$, and $\htheta_\tau= 1$, otherwise.
Similar to the scalar system, the attacker learns in the exploration phase only, and uses the LS learning algorithm, which implies that
 \begin{align}
\label{learningAlgorithmvect-var}
	\hA_{L}=     
	\begin{cases}
        0_{n \times n}, & \mbox{det}(G_{L-1})=0;
	 \\ 
	 \sum\limits_{k=1}^{L-1}(X_{k+1}-U_k)X^{\top}_k G_{L-1}^{-1}, & \textrm{otherwise} ,
    \end{cases} 
\end{align}
where $G_\tau\triangleq\sum_{k=1}^{\tau} X_k X_k^{\top}$. Our simulation parameters are the following:
$\gamma=0.1$, $A=[1~~2~; ~3~~ 4 ]$, $\Sigma=[1~~0~; ~0~~ 1 ]$, and $U_k = -0.9A Y_k$.

\figref{fig:avalsimu} compares 
the attacker's success rate, as a function of  sizes of detection window $\tau$ for different duration $L$  of the exploration phase.  
The false-alarm rate decreases to zero as the duration of the $\tau$ detection window tends to infinity, similarly to the argument for scalar systems. Thus, as the size of the detection window grows, the success rate of learning-based attacks increases. Finally, as as seen in~\figref{fig:avalsimu}, as the duration of the exploration phase $L$ increases, the attacker's success rate increases, since the attacker improves its estimate of $A$ as $L$ increases. This is in line with the theoretical findings in Theorem \ref{Ltheorem11}.

\vspace{-0.09cm}
 \section{Conclusions and Future Directions}\label{sec:conc}

We have presented tight lower and upper bounds on the expected deception time for learning based MITM attacks, as the probability of correct detection tends to one. Additionally, we provided an order-optimal characterization of the length of the attacker's exploration phase and computed a lower bound on the control cost. 
In the future, we plan to study  online phase learning based attacks, where  the attacker can choose to switch between exploration and exploitation phases dynamically. We also plan to study   methods to mitigate these attacks and render the system secure. The extension of our  results to partially-observable linear vector systems where the input (actuation) gain is unknown, and the characterization of securable and unsecurable subspaces, similar to \cite{satchidanandan2018control},   is another possible research direction. Further extensions to nonlinear systems are also of interest. 


\acks{
We gratefully acknowledge support from NSF awards CNS-1446891 and ECCS-1917177.
}

\bibliography{output} 

\newpage
\appendix

\section{Proof of Proposition \ref{lemma:KLDivergence}}
\begin{proof}
Since the attacker does not intervene before $k\leq L$,  we have that for all $k\leq L$,
\begin{align}
    D(p_{1}(Y_{1:k})||p_{0}(Y_{1:k}))=0.
\end{align}
Thus, for all $k > L$, using the chain rule, we have
\begin{align}
\label{feiej123}
D(p_{1}(Y_{1:n})||p_{0}(Y_{1:n}))=
&\sum_{k=L+1}^{n}D(p_{1}(Y_{k}|Y_{1:k-1})||p_{0}(Y_{k}|Y_{1:k-1})).
\end{align}
 Also, if $\Theta_{k}=1$, then for all $k>L$, we have 
\begin{equation}\label{eq:distribut1}
    Y_{k}|(Y_{k-1},U_{k-1},\Hat{A}_{k-1}) \sim \mathcal{N}(\Hat{A}_{k-1}Y_{k-1}+U_{k-1},\sigma^2 I_M),
\end{equation}
since $Y_k=V_k$ for all $k>L$.
Similarly, if $\Theta_{k}=0$, then for all $k>L$, we have
\begin{equation}\label{eq:distribut2}
    Y_{k}|(Y_{k-1},U_{k-1},\Hat{A}_{k-1} )\sim\mathcal{N}(AY_{k-1}+U_{k-1},\sigma^2 I_M).
\end{equation}
The result now follows by using the fact that for all $k>L$, we have $Y_k=V_k$ .

We continue by noticing that the control input $U_{k}$ lies in sigma field of past observations, namely $U_{k}$ is measurable with respect to sigma field generated by $Y_{1:k-1}$. Thus, combining \eqref{feiej123},  \eqref{eq:distribut1} and \eqref{eq:distribut2}, 
for all $k>L$, we have that
\begin{equation}
\label{ejifs1234!}
\begin{split}
     D(p_{1}(Y_{k}|Y_{1:k-1})||p_{0}(Y_{k}|Y_{1:k-1}))
     &=\mathbb{E}\bigg[\frac{Y_{k-1}^{T}(\Hat{A}_{k-1}-A)^T(\Hat{A}_{k-1}-A)Y_{k-1}}{2\sigma^2}\bigg].
\end{split}
\end{equation}
Using ~\eqref{feiej123} and~\eqref{ejifs1234!}, for all $n>L$, we have
\begin{equation}
\begin{split}
    D(p_{1}(Y_{1:n})||p_{0}(Y_{1:n}))
    &=\mathbb{E}\bigg[\sum_{k=L+1}^{n}\frac{Y_{k-1}^{T}(\Hat{A}_{k-1}-A)^T(\Hat{A}_{k-1}-A)Y_{k-1}}{2\sigma^2}\bigg].
\end{split}
\end{equation}
\end{proof}
\section{Proof of the Theorem \ref{thm:timeErrorEnergy}}
\begin{proof}

The proof of the theorem consists of two parts.  First, for all  attacks in the class $\mathcal{A}{(L)}$ and $0<c<1$, we show that  if the probability of error 
is small, namely $\mathbb{P}
    ( \htheta_\tau\neq \Theta_\tau)=
        O(|\epsilon\log\epsilon|)$, then the log-likelihood ratio $\log(p_{1}(y_{1:\tau})/p_{0}(y_{1:\tau}))$ should be greater than $(1-c)\log (1/ \epsilon)$ with high probability  as $\epsilon\to 0$, namely
\begin{equation}\label{eq:2}
     \log\frac{p_{1}(y_{1:\tau})}{p_{0}(y_{1:\tau})}\geq (1-c)\log (1/\epsilon)
\end{equation}
must hold with high probability, as $\epsilon\to 0$. Second, we show that there exists  $0<\bar{c}<1$ such that for all $0<c\leq \bar{c}$ and $T(\epsilon)<(1-c) \log  (1/\epsilon) /C(n_0)$, it is unlikely that the inequality in \eqref{eq:2} is satisfied. 

 Using \eqref{eq:Error1}, for all $k\geq L+1$, we have that
 both type $\RN{1}$ and type $\RN{2}$ errors of the hypothesis test $\Theta_{k}=1$ vs. $\Theta_{k}=0$ are $O(|\epsilon\log \epsilon| )$. Thus, using \cite[Lemma 4]{chernoff1959sequential}, for all $0<c<1$, we have
\begin{equation}
\label{eq:1}
\mathbb{P}\Big(S^{\tau}\leq -(1-c)\log \epsilon \Big) = O(-\epsilon^{c}\log \epsilon ),
\end{equation}
where 
\begin{equation}
    S^n=\log \frac{p_{1}(y_{1:n})}{p_{0}(y_{1:n})}=\sum_{k=1}^{n} \log\bigg(\frac{p_{1}(y_{k}|y_{1:k-1})}{p_{0}(y_{k}|y_{1:k-1})}\bigg). 
\end{equation}
Therefore, as $\epsilon\to 0$, the probability in \eqref{eq:1} tends to 0, which concludes the first part of the proof.

Now, we show that for all $0<c<1$, we have
\begin{equation}
\label{eq:3}
\lim_{n^{\prime} \rightarrow \infty} \mathbb{P}\left(\max_{1\leq k \leq n^{\prime}}S^{k} \geq (D(p_{1}(y_{1:n^{\prime}})||p_{0}(y_{1:n^{\prime}}))+n^{\prime}c)\right)= 0,
\end{equation}
where $D(p_1(y_{1:n^\prime})||p_0(y_{1:n^\prime}))$ denotes the KL divergence between the distributions $p_{1}$ and $p_0$ of $Y_{1:n^\prime}$. We have 
\begin{equation}
\begin{split}
S^{n} &= \sum_{k=1}^{n} \Bigg( \log\bigg(\frac{p_{1}(y_{k}|y_{1:k-1})}{p_{0}(y_{k}|y_{1:k-1})}\bigg)-D(p_{1}(Y_{k}|Y_{1:k-1})||p_{0}(Y_{k}|Y_{1:k-1})) \Bigg)\\
&\quad+\sum_{k=1}^{n}D(p_{1}(Y_{k}|Y_{1:k-1})||p_{0}(Y_{k}|Y_{1:k-1}))\\
&=M_{1}^{n}+M_{2}^{n},
\end{split}
\end{equation}
where
\begin{equation}
\begin{split}
M_{1}^{n}&= \sum_{k=1}^{n} \Bigg( \log\bigg(\frac{p_{1}(y_{k}|y_{1:k-1})}{p_{0}(y_{k}|y_{1:k-1})}\bigg)-D(p_{1}(Y_{k}|Y_{1:k-1})||p_{0}(Y_{k}|Y_{1:k-1})) \Bigg),\\
\end{split}
\end{equation}
is a martingale with mean 0 with respect to filtration $\mathcal{F}_{k}=\sigma(Y_{1:k-1})$, and 
\begin{equation}
\begin{split}
    M_{2}^{n}&=\sum_{k=1}^{n}D(p_{1}(Y_{k}|Y_{1:k-1})||p_{0}(Y_{k}|Y_{1:k-1})),\\
    &\stackrel{(a)}{=}D(p_{1}(Y_{1:n})||p_{0}(Y_{1:n})),
\end{split}
\end{equation}
where $(a)$ follows from the chain rule of KL-Divergence. Now, if the event in \eqref{eq:3} occurs for a fixed $n_1$, namely
\begin{equation}
    M_{1}^{n_1}+M_{2}^{n_1}\geq D(p_{1}(Y_{1:n_1})||p_{0}(Y_{1:n_1}))+n_{1}c,
\end{equation}
then it implies that $M_{1}^{n_1}\geq n_{1}c.$  
Since $Y_{k}|Y_{1:k-1}$ has a normal distribution using  \eqref{eq:distribut1} and \eqref{eq:distribut2}, there exists a constant $b>0$ such that the probability in \eqref{eq:3} simplifies as 
\begin{equation}
\begin{split}
 \mathbb{P}\left(\max_{1\leq k \leq n^{\prime}}S^{k} \geq (D(p_{1}(y_{1:n^{\prime}})||p_{0}(y_{1:n^{\prime}}))+n^{\prime}c)\right)
 &\leq \mathbb{P}(\max_{1\leq k\leq n^{\prime}} M_{1}^{k}\geq n^{\prime}c) \stackrel{(a)}{\leq} {b}/{n^{\prime}c^2},
\end{split}
\end{equation}
where $(a)$ follows from the Doob-Kolmogorov extension of Chebyshev's 
inequality in \cite{doob1953stochastic}, and the fact that $M_{1}^{k}$ is a martingale with 0 mean. Hence, we have that \eqref{eq:3} follows.

Now, we have
\begin{equation}
    n_0C(n_0)< \log(1/\epsilon).
\end{equation}
Therefore, there exists $0<\bar{c}<1$ such that 
\begin{equation}
    n_0 C(n_0)+n_{0}\bar{c}=(1-\bar{c}) \log(1/\epsilon).
\end{equation}
Now, using Proposition \ref{lemma:KLDivergence}, for all $0<c\leq \bar{c}$, we have
\begin{equation}
\begin{split}
\mathbb{P}(N\leq n_{0})
&\leq \mathbb{P}\Big( N \leq n_{0} \mbox{ and } S^{N}\geq n_{0}(C(n_{0})+{c}) \Big)+\mathbb{P}\Big(
  S^{N}
\leq n_{0}(C(n_{0})+{c})\Big)\\
&\leq \mathbb{P}\Big( \max_{1\leq k \leq n_{0}} S^{k}\geq n_{0}(C( n_{0})+{c})\Big)+\mathbb{P}\Big(
  S^{N}
\leq  n_{0}(C( n_{0})+{c})\Big),
\end{split}
\label{eq:5}
\end{equation}
and the first and the second terms at the right-hand side of \eqref{eq:5} approach   zero  by  \eqref{eq:3} and \eqref{eq:1},  respectively.
\end{proof}

\section{Proof of the Theorem \ref{thm:ExistenceOfTest}}
\begin{proof}
In $\mathcal{A}(L)$, consider an  attack $R^*$  such that for all $k>L$, we have $  \hA_k=\hA_L. $
For all $k>L$, if $\Theta_{k}=1$, then we have 
\begin{equation}
    Y_{k}|Y_{1:k-1} \sim \mathcal{N}(\Hat{A}_{L}Y_{k-1}+U_{k-1},\sigma^2 I_M).
\end{equation}
Similarly, if $\Theta_{k}=0$, then
\begin{equation}
    Y_{k}|Y_{1:k-1}\sim\mathcal{N}(AY_{k-1}+U_{k-1},\sigma^2 I_M).
\end{equation}

Consider a the following detection strategy, also known as Sequential Probability Ratio Test (SPRT), at the controller as follows. At time $n$, if 
\begin{equation}\label{eq:con1}
    \sum_{k=1}^{n} \log\bigg(\frac{p_{1}(y_{k}|y_{1:k-1})}{p_{0}(y_{k}|y_{1:k-1})}\bigg)\geq \log(1/\epsilon),
\end{equation}
then $\hat{\Theta}_{n}=1$, and if 
\begin{equation}\label{eq:con2}
     \sum_{k=1}^{n} \log\bigg(\frac{p_{0}(y_{k}|y_{1:k-1})}{p_{1}(y_{k}|y_{1:k-1})}\bigg)\geq \log(1/\epsilon),
\end{equation}
then $\hat{\Theta}_{n}=0$. Otherwise, $n$ is not a decision time and the test continues. 

We will show that for the attack $R^*$ and the detection strategy SPRT, the statement of the theorem holds. 

For SPRT, the probability of error, both $\Pdec^{\tau}$ and $P_{FA}^{\tau}$, is at most $O(\epsilon)$, and the proof is along the same direction as \cite[Theorem 1]{rangi2018distributed}. Now, let us prove the bound on $T(\epsilon)$. Given the system is under attack, let the decision time $\tau$ of SPRT be
\begin{equation}
T=\min\bigg\{n:\sum_{k=1}^{n} \log\bigg(\frac{p_{1}(y_{k}|y_{1:k-1})}{p_{0}(y_{k}|y_{1:k-1})}\bigg)\geq \log(1/\epsilon)\bigg\}.
\end{equation}
Using \cite[Lemma 2]{chernoff1959sequential}, for system under attack $\mathcal{A}(L)$ and for all $c>0$, there exist a $b>0$ such that
\begin{equation}\label{eq:KLDivergence1}
\begin{split}
     \mathbb{P}\bigg(\sum_{k=1}^{n}\log\bigg(\frac{p_{1}(y_{k}|y_{1:k-1})}{p_{0}(y_{k}|y_{1:k-1})}\bigg)<(D(p_{1}(Y_{1:n})||p_{0}(Y_{1:n}))-nc)\bigg)\leq e^{-bn}.
\end{split}
\end{equation}
Using the definition of $n_0$, for all $\bar{n} > n_0$ we have
\begin{equation}\label{eq:KLDivergence2}
\begin{split}
     \log(1/\epsilon)&\leq \bar{n}C(\bar{n})= 
     D(p_{1}(Y_{1:\bar{n}})||p_{0}(Y_{1:\bar{n}})),
\end{split}
\end{equation}
where the equality follows from Proposition \ref{lemma:KLDivergence}. 
Using \eqref{eq:KLDivergence1} and \eqref{eq:KLDivergence2}, 
For all $c>0$ and $n\geq (1+c)(n_0+1)\log(1/\epsilon)/D(p_{1}(Y_{1:n_0+1})||p_{0}(Y_{1:n_0+1}))$, we have
\begin{equation}
\begin{split}
     &\mathbb{P}\bigg(\sum_{k=1}^{n}\log\bigg(\frac{p_{1}(y_{k}|y_{1:k-1})}{p_{0}(y_{k}|y_{1:k-1})}\bigg)< \log(1/\epsilon)\bigg)\leq e^{-bn}.
\end{split}
\end{equation}
Then, using Proposition \ref{lemma:KLDivergence}, the statement of the theorem follows. 
\end{proof}

\section{Proof of Theorem \ref{Ltheorem11}}
\begin{proof} 
If the learning algorithm  in the exploration phase is a convergent algorithm,  the learning algorithm  in the exploitation phase is a non-divergent algorithm, then for all $0<\delta<1$, we have
\begin{equation}\label{eq:energyFictSignal}
\begin{split}
    C(n_0)&\stackrel{(a)}{\leq} \|\hA_{L}-A\|_{op}^2 \frac{1}{n_0} \mathbb{E}\bigg[\sum_{k=L+1}^{n_0}\frac{V_{k-1}^{T}V_{k-1}}{2\sigma^2}\bigg],\\
    &\stackrel{(b)}{\leq} \bigg(\frac{c^{1/\alpha}}{L\delta^{1/\alpha}}\bigg)\tilde{C}(n_0),
\end{split}
\end{equation}
with probability at least $1-\delta$, where $(a)$ follows from the fact that
\begin{equation}
    ||Ax||_2\leq ||A||_{op}||x||_2 ,
\end{equation}
and the  learning algorithm in the exploitation phase is non-divergent, 
and $(b)$ follows from Definition \ref{definitofLeanif} of convergent algorithms. Thus, we have 
\begin{equation}\label{eq:lowBoundKL}
    \frac{\log(1/\epsilon)}{C(n_0)}\geq \frac{\log(1/\epsilon)}{\tilde{C}(n_0)} \bigg(\frac{L\delta^{1/\alpha}}{c^{1/\alpha}}\bigg),
\end{equation}
with probability at least $1-\delta$. 
Using Theorem \ref{thm:timeErrorEnergy} and \eqref{eq:lowBoundKL}, if
\begin{equation}
   (1+o(1))\frac{\log(1/\epsilon)}{\tilde{C}(n_0)} \bigg(\frac{L\delta^{1/\alpha}}{c^{1/\alpha}}\bigg)>D(1+o(1)), \mbox{ as }\epsilon\to 0, 
\end{equation}
then $\mathbb{E}[T(\epsilon)]> D+o(1)$ as $\epsilon\to 0$. This along with \eqref{eq:lowBoundKL} implies that
\begin{equation}
L\geq {\frac{(1+o(1)) D \tilde{C}(n_0)}{\log(1/\epsilon)}}\frac{c^{1/\alpha}}{\delta^{1/\alpha}}, \mbox{ as }\epsilon\to 0, 
\end{equation}
with probability at least $1-\delta$. 
\end{proof}
\section{Proof of Theorem \ref{thm:algExistence}}
\begin{proof}
 Consider the LS learning algorithm in \cite{simchowitz2018learning} which satisfies 
 \begin{align}
    \mathbb{P}(\|\hA_{k}-A\|_{op}>\eta) \le \frac{c_1}{e^{\eta^2 k}},
\end{align}
For $\eta=\sqrt{\log(c_1/\delta)/L}$, similar to \eqref{eq:energyFictSignal}, we have that 
\begin{equation}
    C(n_0)\leq \frac{\log(c_1/\delta)}{L}\tilde{C}(n_0),
\end{equation}
with probability at least $1-\delta$. Thus, we have 
\begin{equation}
    \frac{\log(1/\epsilon)}{C(n_0)}\geq \frac{\log(1/\epsilon)}{\tilde C(n_0)} \frac{L}{\log(c_1/\delta)},
\end{equation}
with probability at least $1-\delta$. Thus, for $L=(1+o(1))D \tilde C(n_0) \log(c_1/\delta) /\log(1/\epsilon)$ as $\epsilon\to 0$, using Theorem \ref{thm:timeErrorEnergy}, we have that
\begin{equation}
    \mathbb{E}[T(\epsilon)]\geq \frac{(1+o(1))\log(1/\epsilon)}{C(n_0)}\geq D(1+o(1))=D+o(1), \mbox{ as }\epsilon\to 0,
\end{equation}
with probability at least $1-\delta$. The statement of the theorem follows. 
\end{proof}

\section{Proof of Theorem \ref{controlenergw22}}
\begin{proof}
 Since $\tW_k$ is independent of $U_k$ and $V_k$ and $\mathbb{E}[\tW_k]=0$, we have 
  \begin{align}
  \label{labelforconditionn!}
      \mathbb{E}[V_{k+1}^TV_{k+1}]-\mathbb{E}[U_{k}^TU_k]=
     \mathbb{E}[V_k^T\hat{A}^T_k \hat{A}_k V_{k}]+\sigma^2+ 2\mathbb{E}[V^T_{k}\hat{A}^T_kU_k].
  \end{align}
Using \eqref{kkdcondition!}, we have
  \begin{align}\label{whatwew1!}
      \mathbb{E}[V_{k+1}^TV_{k+1}] \leq \mathbb{E}[U_{k}^TU_{k}], 
  \end{align}
which implies 
\begin{equation}\label{eq:energyBound1}
\begin{split}
     C(n_0)&\stackrel{(a)}{\leq} \frac{\|\hA_{L}-A\|_{op}^2}{n_0} \mathbb{E}\bigg[\sum_{k=L+1}^{n_0}\frac{V_{k-1}^{T}V_{k-1}}{2\sigma^2}\bigg]\\
     &\stackrel{(b)}{\leq} \frac{\|\hA_{L}-A\|_{op}^2}{n_0} \mathbb{E}\bigg[\sum_{k=L}^{n_0-1}\frac{U_{k-1}^{T}U_{k-1}}{2\sigma^2}\bigg],
\end{split}
\end{equation}
where $(a)$ follows from the fact that $ ||Ax||_2\leq ||A||_{op}||x||_2 $, and $(b)$ follows from \eqref{whatwew1!}. 
Since $\mathbb{E}[T(\epsilon)]\leq D+o(1)$ as $\epsilon\to 0$, using Theorem \ref{thm:timeErrorEnergy} and \eqref{eq:energyBound1}, we have that
\begin{equation}
    D+o(1)\geq \frac{(1+o(1))2\sigma^2\log(1/\epsilon)}{\|\hA_{L}-A\|_{op}^2 R(n_0)}, \mbox{ as }\epsilon\to 0.
\end{equation}
Hence, the statement of the theorem follows.
\end{proof}

\end{document}